\documentstyle[epsfig]{l-aa}
\def\jref#1 #2 #3 #4 {{\par\noindent \hangindent=2em \hangafter=1
      \advance \rightskip by 0em #1, {\it#2}, {\bf#3}, #4.\par}}
\def\rref#1{{\par\noindent \hangindent=2em \hangafter=1
    \advance \rightskip by 0em #1.\par}}

\def\rosat{{\it ROSAT}}
\def\sax{{\it BeppoSAX}}
\def\ulysses{{\it Ulysses}}
\def\wind{{\it Wind}}
\def\rerg{\rm erg}
\def\rs{\rm s}
\def\rs1{\rm s^{-1}}

\def\rcm{\rm cm}
\def\rcm2{\rm cm^{-2}}

\def\flux{\rerg\ \rcm2\ \rs1}

\def\etal{{\it et al. }}

\begin{document}

\title{High resolution imaging of the X-ray afterglow of GRB970228 with \rosat}
\author{
F. Frontera\inst{1,2}
\and J. Greiner\inst{3}
\and L.A. Antonelli\inst{4}
\and E. Costa\inst{5}
\and F. Fiore\inst{4}
\and A.N. Parmar\inst{6}
\and L. Piro\inst{5}
\and T. Boller\inst{7}
\and W. Voges\inst{7}
}

\thesaurus{13
           (13.07.1;   
            13.25.1;)} 
\institute{
{Istituto Tecnologie e Studio Radiazioni Extraterrestri, CNR, Via 
Gobetti, 101, 40129 Bologna, Italy}
\and
{Dipartimento di Fisica, Universit\`a di Ferrara, Via Paradiso, 11,
44100 Ferrara, Italy}
\and
{Astrophysikalisches Institut Potsdam, 14482 Potsdam, Germany}
\and
{BeppoSAX Scientific Data Centre, c/o Nuova Telespazio, Via Corcolle 19,
00131 Roma, Italy}
\and
{Istituto di Astrofisica Spaziale, CNR, Via E. Fermi, 00044 Frascati, Italy}
\and
{Astrophysics Division, Space Science Department of ESA, ESTEC, NL-2200 AG
Noordwijk, The Netherlands}
\and
{Max-Planck-Institute fuer Extraterrestrische Physik, 85740 Garching,
Germany}
}

   \offprints{F. Frontera:filippo@tesre.bo.cnr.it}
   \date{Received ....... / Accepted ...........}
    \maketitle
 \markboth{F. Frontera et al.}{GRB970228 with {\it ROSAT}}

\begin{abstract}
We report results of a \rosat\ High-Resolution Imager
(HRI) observation of the X-ray error box
given by the \sax\ Wide Field Camera for the gamma-ray burst 
that occurred on 1997 February 28.
The observation started 10 days after the burst and ended
three days later, with a total exposure of 34.3~ks.
An X-ray source was detected within the 3$'$ WFC error box and its
position determined with a 10$''$ radius accuracy. The source position is
in the \sax\ Narrow Field Instrument source error box and
is coincident (to within 2$''$) with the optical transient associated 
with GRB970228.
This is the most precise position obtained for an X-ray afterglow and 
confirms that the X-ray and optical afterglows have the same origin.
We present the 0.1--2.4~keV combined HRI and \sax\ Low-Energy Concentrator
Spectrometer decay light curve which can be well fit with a power-law.  
The decay is consistent with
that measured at higher energies (2--10~keV) with the \sax\ Medium-Energy
Concentrator Spectrometer.


\end{abstract}

\section{Introduction}

Observations of celestial Gamma-Ray Bursts (GRB) 
performed over the last 25 years had not, until recently, succeeded in finding 
counterparts in other wavelength bands. The ability of the
\sax\ satellite to provide arc minute precision 
positions (Piro et al. 1998) and to observe these positions within 
hours of the GRB changed this 
situation in 1997 when the X-ray afterglow of GRB970228 was
measured (Costa et al 1997a). The burst was detected (Costa et al. 1997a)
 with the Gamma-Ray 
Burst Monitor (GRBM) (40--70~keV, Frontera et al. 1997a) on 1997 
February 28.123620 UT and also detected 
in the 1.5--26 keV energy range by one of the two Wide Field Cameras (WFC No. 
1) aboard the same satellite (Jager et al. 1997). Its position was 
determined with an error circle of 3~arcmin (3$\sigma$) radius, centered on
 $\alpha_{2000}\,=\,05^h01^m57^s$, $\delta_{2000}\,=\,11^\circ46'24''$.
Eight hours after the GRB trigger, from February 28.4681 to February 28.8330
UTC, the Narrow Field Instruments (NFI) on board \sax\ (Boella et al. 1997a) 
were pointed to the WFC error box.
An X-ray source, SAX J0501.7+1146, was detected (Costa et al 1997b) in the
field of view of both the Low Energy (0.1--10~keV) and Medium Energy
(2--10~keV) Concentrators Spectrometers (LECS and MECS) (Parmar et al. 1997;
Boella et al. 1997b). The source position ($\alpha_{2000}\,=\,05^h01^m44^s$,
$\delta_{2000}\,=\,11^\circ46'42''$) is consistent with the GRB error
circle. The source was again observed about three days later,
from March 3.7345 to March 4.1174. During this observation, the 2--10~keV
source flux had decreased by about a factor 20, while in the 0.1--2~keV 
energy range the source was not detected.
Following the discovery of the GRB, searches for radio and optical 
counterparts to
GRB970228 were conducted with most of the ground based telescopes in 
the northern hemisphere. Groot et al. (1997) reported the discovery
of an optical transient at a position ($\alpha_{2000}\,=\,05^h01^m46.70^s$,
$\delta_{2000}\,=\,11^\circ46'53.0''$), consistent with both the \sax\ WFC
and NFI error boxes and with the long baseline timing 
\ulysses/\sax\ and  \ulysses/\wind\ error
annuli, of 31$''$ and 30$''$ half-width, respectively 
(Hurley et al. 1997; Cline et al. 1997).
While the association of the transient X-ray source with the
afterglow of GRB970228 was compelling on the basis of the properties of
its decay curve when extrapolated backwards to the burst time (Costa et al.
1997c), the association of the optical transient with the burst afterglow
was less strong.
In spite of the positional consistency and temporal behaviour of the
optical transient, 
it was not possible to exclude the possibility that the optical 
transient was 
unrelated to the GRB (see discussion by van Paradijs et al. 1997), like in
the case of the
radio source discovered in the earliest
error box of GRB970111, which showed  a time behaviour consistent with 
that expected from radio afterglows  of GRBs, but later resulted to be 
unrelated tho the burst  (Frail et al. 1997). 
The \rosat\ satellite, thanks to its HRI focal plane instrument, offered
the possibility of imaging the X-ray afterglow at 10$''$ angular
resolution (David et al. 1997). 
A Target of Opportunity observation was thus requested and
obtained. Here we report on results of that observation and its consequences.
Preliminary results were already previously reported (Frontera et al. 1997b).

\section{Observation and analysis}
The \rosat\ observation started on 1997 March 10 at 18:54:31 UT and ended on
March 13 at 07:41:00 UT with a total exposure time of 34.3~ks. The pointing
coordinates were $\alpha_{2000}\,=\,05^h01^m52.80^s$ and
$\delta_{2000}\,=\,11^\circ46'24''$.
In the telescope field of view of 20$'$ radius, eight sources were detected
at $\geq$3$\sigma$ level in the 0.1--2.4~keV energy band.
Of them, one source, RX J050146+1146.9, was found in the 
3$'$ radius error circle given by the \sax\ WFC for GRB970228.
The other X--ray sources are more than 5$'$ away from the nominal GRB position.
The HRI source position, determined with an error radius of 10$''$ (6$\sigma$
confidence level),
is centered on $\alpha_{2000}\,=\,05^h01^m46.6^s$ and $\delta_{2000}\,=
\,11^\circ46'52.3''$.
Figure 1 shows a part of the HRI field of view with superposed the error
boxes given by the \sax\ WFC and NFIs and the error annulus obtained using
the GRB arrival time technique to the \sax\ GRBM and \ulysses\ detectors
(Hurley et al. 1997).
The plot is a likelihood image which has been obtained by fitting the point
spread function of the HRI at a given off-axis angle to the photon source
on a grid with 2.5$''$ spacing in Right Ascension and Declination (see
Greiner et al. (1995) for a more detailed description of the procedure).
Since we deal only with the central part of the HRI detector, 
we have ignored
all photons below HRI channel 2 and above HRI channel 12 in order to
improve the signal to noise ratio (David et al. 1997).
There is only one source with a likelihood larger than 8 (corresponding to
3~$\sigma$), all other structures in the image are below the 3$\sigma$
significance level.
As can be seen,
the new HRI source is completely inside the larger error circle obtained 
with the \sax\ LECS and MECS instruments and is coincident with the optical 
transient within 2~$''$. 
Using the logN--logS curve obtained with \rosat\ (Hasinger 1998), 
at the sensitivity level achieved in our observation the probability that 
the \rosat\ source is inside the NFI error box (50$''$ radius) by chance 
coincidence is $\sim 1 \times 10^{-3}$.
This shows that the \sax\ transient and the \rosat\ source are likely 
the same object. In addition 
the positional coincidence of the \rosat\ source with the optical 
transient confirms the identification of the optical transient with 
the X-ray transient.
%
%
\begin{figure}
\epsfig{figure=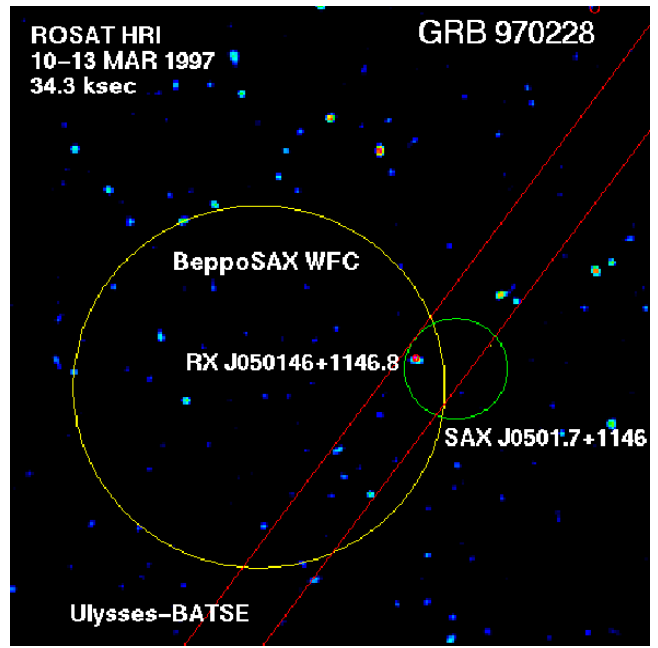,height=8.5cm,width=8.5cm,angle=-90}
\caption{The central 8$'$ of the HRI field of view of the
\rosat\ observation on 1997 March 10--13. This
likelihood image is shown with cut values such that ''sources'' above
1$\sigma$ are visible. The only source above $3\sigma$ in this image 
is RX J050146+1146.8 inside the smaller circle, that is the $\approx$1$'$
error circle of the fading source SAX J0501.7+1146 as found with the two
NFI pointings. The large circle shows the 3$\sigma$ error circle of GRB970228
as determined with the \sax\ WFC and the two straight lines mark the 
triangulation circle derived from the \sax\ and \ulysses\ timings of 
GRB970228 (Hurley et al. 1997).
}
\label{figure:fig1}
\end{figure}

The source flux was derived using intervals when the background level 
was lowest. With this
constraint the useful observing time is reduced to 
10.4~ks (5.3~ks at the beginning
of the observation, 4.1~ks after 60~ks of elapsed time, and 1.0~ks just before
the end of the observation). The count excess due to the source is (1.0$\pm
0.3)\times10^{-3}\, {\rm counts \, s^{-1}}$.
In order to derive the 0.1--2.4~keV source flux  
we assume the spectral shape
measured for the transient source detected with LECS and MECS
during the first \sax\ observation of the WFC error box 
(Frontera et al. 1997c).
This assumption is reasonable, taking into account that the spectral hardness
of the \sax\ source did not appear to change from the first to the second
\sax\ observation (Frontera et al. 1997c).
The spectrum during the first \sax\ pointing is consistent with a
power-law with photon index
$\alpha\,=\,2.1\pm0.3$ and a hydrogen column density of 3.5$^{+3.3}_{-2.3}
\times 10^{21}\, cm^{-2}$ (90\% confidence single parameter errors).
The latter parameter is consistent with the galactic absorption
(N$_H\,=\, 1.6\times10^{21}\, cm^{-2}$) along the source direction.
Assuming the estimated column density,
we find an absorbed 0.1--2.4~keV flux of ($4.0\pm1.5)\times
10^{-14}\, \flux$, and an unabsorbed flux of ($1.6\pm0.6)\times 10^{-13}\,
\flux$ in the same energy range.
The unabsorbed flux is lower than the \rosat\ all-sky survey 
1.7$\sigma$ upper limit in the BSAX/WFC error box of GRB970228
(about 9.2$\times10^{-13}\, \flux$, assuming a 
galactic absorption) (Boller et al. 1997). 
Thus, there is no evidence of variability from the \rosat\ data alone.

\section {Discussion}
An important issue in the study of GRB afterglows is the shape of their
flux decay with time after the initial event as a function of wavelength.
Simple versions of fireball models (e.g., M\'esz\'aros \& Rees 1997) predict 
that the
afterglow decline law is independent of photon energy. Costa et al. (1997c)
report on the decay curve of the X-ray afterglow of GRB970228 in the
2--10~keV energy range. They find that the decay is consistent with a
power-law (t$^{-\alpha}$), where t is the time (in seconds) from the
burst onset and $\alpha \,=\,1.33^{+0.13}_{-0.11}$, for
at least 6.9 days after the initial event.
Galama et al. (1997) reported a more complex time behavior for the 
R-band flux of the optical transient associated with GRB970228.
For about 6 days from the burst the decay could be approximated by 
a power-law with a much higher slope ($\alpha \,=\,2.1^{+0.3}_{-0.5}$)
than found in the 2--10~keV energy range, whereas after 6 
days $\alpha \leq 0.35$. Fruchter et al. (1997), on the basis 
of a $Hubble$ Space Telescope
observation of the same source performed six months after the initial
event (4 September 1997),  found that the optical transient continues
to decline according to a power-law with index $\alpha\,\sim 1.1$, with 
the exception of time period from March 6 to March 13, which 
determined the result reported by Galama et al.(1997).
By combining the \rosat\ observation with the \sax\ observations, we can study
the light curve behavior of the GRB970228 afterglow in a lower
energy band (0.1--2.4~keV) and for a longer time (13 days from the
initial event) than the observations quoted by Costa et al. (1997c).
Figure 2 shows the overall light curve of the source in the 0.1--2.4~keV
energy range
without any correction for photoelectric absorption.
The data points of the first \sax\ observation are those obtained with 
the LECS,
while those of the second observation include both
the LECS 3$\sigma$ upper limit and the extrapolation to the 0.1--2.4~keV band
of the MECS flux measured in the 2--10~keV band, assuming
the same spectrum as measured during the first observation.  
The source decline is  fit with a power-law, $At^{-\alpha}$ (t in seconds), 
with index $\alpha\,=\,1.50^{+0.23}_{-0.35}$ and 
$A\,=\,(3.3^{+27}_{-2.6})\times 10^7$ (in units of $10^{-12} \, \flux$).
Uncertainties are single parameter errors at 90\% confidence level.
This power-law index is fully consistent with that derived by Costa et al.
(1997c) for the contiguous 2--10~keV energy band.
No evidence for a cut-off, as predicted by relativistically expanding 
fireball models
when the GRB remnant becomes non relativistic (Wijers et al. 1997), or
a slower decline in the flux as observed by Galama et al. (1997) for the
same afterglow in the optical band, are evident.
The index of the decline power law is marginally consistent with the 
optical decline slope reported by
Fruchter et al. (1997, 1998).  The backward extrapolation of the best-fit
power-law to the time of the burst (Fig.~3) gives a flux of
of 8.4$^{+68}_{-7.2}\times 10^{-8}\, \flux$ in the burst time interval
35--80~s from its onset,
when the X-ray afterglow is likely to start (Costa et al. 1997c, 
Frontera et al. 1997c).
This value, when compared with the \sax/ WFC average 2--10~keV flux, 
in the same burst time interval, of 3.6$^{+0.7}_{-0.6}\times 10^{-8}\, \flux$,
is consistent with the extrapolation of the energy spectrum of the burst 
in the same interval (a power-law with a photon index of 
$\sim 1.6$ for E$\ge$2~keV, Frontera et al. 1997c).
%
%
\begin{figure}
\epsfig{figure=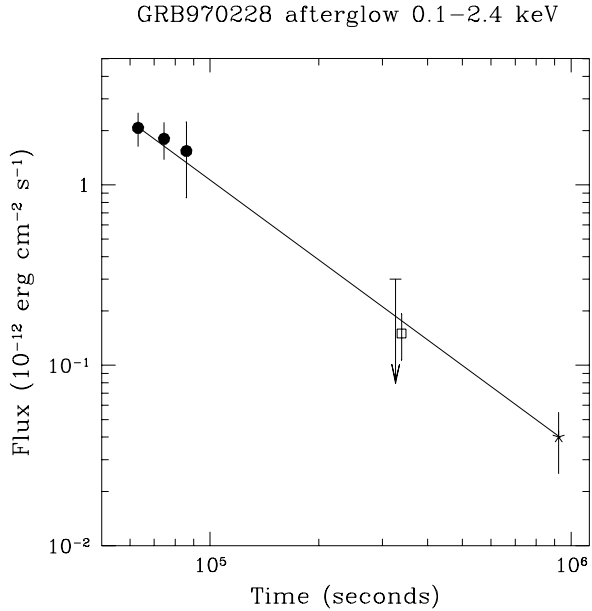,height=8.5cm,width=8.5cm,angle=0}
\caption{ 
The decline of the 0.1--2.4~keV source flux with time from the burst
onset, uncorrected for galactic absorption. The filled dots are the LECS
data points, the arrow is the LECS 3$\sigma$ upper limit,
the square gives the flux extrapolated
from the MECS detection (see text) and the cross shows
the \rosat\ HRI data point. 
The best fit power-law decay is also shown.}
\label{figure:decaya}
\end{figure}

\section{Conclusions}
The \rosat\ HRI observation of the \sax\ WFC error circle of GRB970228 
clearly shows the presence of a new X-ray source.
Its position  within the error box of the \sax\ source, the low probability
of a chance coincidence ($\sim 1 \times 10^{-3}$) and 
the better imaging capabilities of the \rosat\ HRI compared to the 
\sax\ NFI, indicate that 
the \rosat\ source and the \sax\ source are the same object. The source
position derived from the \rosat\ observation is the most precise position
of a GRB X-ray afterglow obtained thus far. Its position is also coincident with
the optical transient associated to GRB970228 within 2$''$. This result
confirms that X-ray source and the optical transient are the same object.  
The X-ray source hows a 0.1--2.4 keV decline according to a power law 
decline with index $\alpha \,=\, 1.50^{+0.23}_{-0.35}$ for at least 13 days. 
This slope is fully consistent with that estimated in the 2--10 keV energy 
band (Costa et al. 1997c) and is marginally consistent with that reported 
by Fruchter et al. (1997, 1998) in the optical band. Thus it appears that
from the X-ray to the optical band the GRB afterglow has the same decline
law.
%
%
\begin{figure}
\epsfig{figure=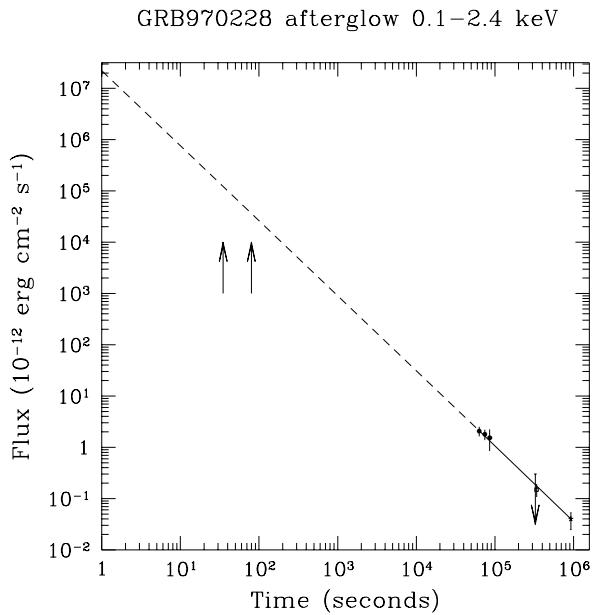,height=8.5cm,width=8.5cm,angle=0}
\caption{As in Fig. 2, but extrapolated to the first second from 
the burst onset. The two arrows on the left delimit the time
interval, that corresponds to the GRB last three pulses, when
the X-ray afterglow is expected to start (see text).
}
\label{figure:decayb}
\end{figure}

\begin{acknowledgements}
It is a great pleasure to thank J. Tr\"umper for granting the ROSAT 
target-of-opportunity time.
JG is supported by the German Bundesministerium f\"ur Bildung,
Wissenschaft und Forschung (BMBF/DLR) under contract No. 50 QQ 9602 3. 
This research was supported by the Italian Space Agency ASI.
The ROSAT project is supported by the BMBF/DLR and the Max-Planck-Society.
\end{acknowledgements}

\end{document}